\documentclass[usletter,prl,twocolumn]{revtex4-1}
\usepackage{graphicx}
\usepackage{amsmath,amsfonts,amssymb}
\usepackage{epstopdf}
\usepackage{hyperref}
\begin{document}
\title{Importance of Reversibility in the Quantum Formalism
}

\author{Fran\c cois David}
\affiliation{Institut de Physique Th\'eorique\\
CNRS, URA 2306, F-91191 Gif-sur-Yvette, France\\
CEA, IPhT, F-91191 Gif-sur-Yvette, France}
\email[]{francois.david@cea.fr}
\thanks{I thank M.-C. David, M. Bauer and V. Pasquier for help and comments.}
\date{October 13, 2011}                                  
\begin{abstract} In this letter I stress the role of causal reversibility (time-symmetry), together with causality and locality, in the justification of the quantum formalism. 
Firstly, in the algebraic quantum  formalism, I show that the assumption of reversibility implies that the observables of a quantum theory form an abstract real C$^\star$\nobreakdash-algebra, and can be represented as an algebra of operators on a real Hilbert space. 
Secondly, in the quantum logic formalism, I emphasize which axioms for the lattice of propositions (existence of an orthocomplementation and the covering property) derive from reversibility. A new argument based on locality and Soler's theorem is used to derive the  representation as projectors on a regular Hilbert space from the general quantum logic formalism.
In both cases it is recalled that the restriction to complex algebras and Hilbert spaces comes from the constraints of locality and separability.
\end{abstract}
\maketitle
The principles of quantum physics date back more than 80 years \cite{Dirac30,vonNeumann32G}, and have proved extraordinary successful and robust, leading to the formalism  of quantum field theory.
Despite these tremendous successes, discussions of the foundations and interpretations of the quantum theory 
are more lively than ever.

The purpose of this letter is to show that the (well known) property of reversibility or time-symmetry of the quantum formalism is in fact a crucial feature (together with causality and locality) for its construction and its justification. 
This will be done for the two main proper formulations of quantum theory, algebraic quantum theory and quantum logic.
I first present a new way to introduce the algebraic quantum formalism, and show that causality and reversibility alone lead to a formulation of quantum mechanics in terms of real algebras of operators on real Hilbert spaces (on $\mathbb{R}$), causality being associated to the associative algebra structure, and reversibility to the transposition/involution or $^*$ operation. I then recall why it is locality/separability that enforces the use of complex Hilbert spaces (on $\mathbb{C}$). 
Secondly I consider the quantum logic approach, where causality is the main ingredient. I show that reversibility enters crucially in the definition of the negation (or orthocomplement) of  quantum propositions, a fact implicit in the standard presentations of this formalism, but which does not seem to have been explicitly stated and fully appreciated before. 
I also give a new simple argument for why locality leads to the standard formulation of quantum mechanics in terms of complex Hilbert spaces in this approach.

The discussion done here does not lead to any change of the (already existing) quantum formalism (as usual). But I think that it sheds new light on it, and will be useful for discussions of its (still often puzzling) aspects and its consequences, as well as when discussing alternates theories and extensions of quantum theory.

Some words on the terminology. 
Firstly, by causality I simply mean that in the physical theories considered a clear distinction between past and future can be made, allowing to separate (in some way) causes from effects, but not excluding the possibility of causal independence (like in special relativity where space-like vectors separate the past cone from the future cone).
Secondly, by reversibility I simply mean that there is a symmetry between past and future, in the sense that the laws of physics take the same form irrespective of the choice of a ``time direction'' (in other word  ``there is no microscopic time arrow'').
In the sense given above, these two properties of causality and reversibility are satisfied both in classical physics, in (at least special-)relativistic physics, and (for many authors) in quantum physics. But at a purely logical level they do not necessarily need to be. For instance in classical deterministic dynamics, a  dissipative dynamics in causal but not reversible, while Hamiltonian dynamics is both causal and reversible. In classical stochastic dynamics, a general Markov process is causal but not reversible; it becomes reversible if it fulfills detailed balance.
Finally, about reversibility, when there is a continuous time evolution one usually speaks of time-symmetry, but since in this letter I shall not discuss dynamical aspects, and I often do not even need to assume there is a continuous time, I shall prefer to use the term ``causal reversibility''.

A priori the arguments presented here are valid for non relativistic quantum mechanics and for relativistic quantum (field) theory. I shall not discuss quantum fields in classical gravitational backgrounds. I do not aim at mathematical rigor, and I shall keep the empirical and pragmatic point of view of most physicists, discussing the consistency of the formalism and its applicability before the questions of interpretation.
Details will be published elsewhere \cite{David2011}.

Let me first discuss the algebraic formalism.
In quantum -- as well as  classical -- physics a system is characterized by the sets of its (mixed) states and of its observables. 
Classically observables are real (smooth) functions on a phase space $\Omega$, and statistical states are probability distributions on $\Omega$. 
The observables thus form a real commutative Poisson algebra $\mathcal{A}$, the states being the real linear positive and normalised forms on $\mathcal{A}$.
In the algebraic quantum formalism (see e.g. the authoritative book \cite{Haag96}) the physical observables 
are assumed to generate an associative but non commutative complex algebra $\mathcal{A}_{\scriptscriptstyle{\mathbb{C}}}$ of operators, endowed with a  complex C$^\star$-algebra structure. This structure allows us to represent $\mathcal{A}_{\scriptscriptstyle{\mathbb{C}}}$ as an algebra of operators on a complex Hilbert space $\mathcal{H}_{\scriptscriptstyle{\mathbb{C}}}$. Quantum mixed states are linear positive forms on $\mathcal{A}_{\scriptscriptstyle{\mathbb{C}}}$.

Here I propose to relax to assumption of a complex structure, and to start from an abstract real associative unital algebra $\mathcal{A}_{\scriptscriptstyle{\mathbb{R}}}$ of observables. 
Formulating quantum mechanics in terms of real Hilbert spaces is of course not a new idea (see \cite{Stueckelberg1960}), but the use of  real abstract algebras seems new. It is natural to start from a real algebra of observables  since generic physical measurements give real outcomes. This is the simplest generalisation of the commutative real algebra of classical observables (functions) on some classical configuration space $X$.

The algebra $\mathcal{A}_{\scriptscriptstyle{\mathbb{R}}}$ is thus first endowed with a real vector space structure. It means that any linear real combinations $\mathbf{c}=\lambda \mathbf{a} +\mu \mathbf{b} $ of observables is still an observable ($\mathbf{a},\,\mathbf{b}\in \mathcal{A}_{\mathbb{R}}$, $ \lambda,\, \mu\in\mathbb{R}$). This is known to be  a  non trivial assumption in the non-commutative case, since it ultimately leads to the superposition principle.

The algebra structure from the  product $\mathbf{c}=\mathbf{ab}$ is related to causality. This is known to be the case in the standard quantum formalism, where the product $\mathbf{ab}$ of two local operators is obtained in the Heisenberg picture as the equal time limit  of  time ordered  operators $\lim_{t\to 0_-}\mathbf{a}(0)\mathbf{b}(t)$ (eventually through some OPE), and also per se in the path integral formalism.
Thus I propose to view the product  $\mathbf{ab}$ as a very abstract ``causal combination'' or ``causal succession'' of ``$\mathbf{a}$ after $\mathbf{b}$'', compatible with the addition law.
But to make this product compatible with the usual concept of ``linear'' causality, it is natural to assume that  causal succession of combinations ($\mathbf{abc}\dots$ meaning ``$\mathbf{a}$ after $\mathbf{b}$ after $\mathbf{c}$, etc. ") is independent of the details of the grouping, $\mathbf{(ab)c}=\mathbf{a(bc)}$. This is nothing but associativity for the algebra $\mathcal{A}_{\scriptscriptstyle{\mathbb{R}}}$.
At that stage a state $\varphi$ is just a  real linear form $\mathbf{a}\to\varphi(\mathbf{a})$ on $\mathcal{A}_{\scriptscriptstyle{\mathbb{R}}}$, which gives the expectation value of the operator $\mathbf{a}$ in the state $\varphi$, with physical constraints to be discussed later.

What does  the assumption of causal reversibility mean in this framework? I expect that to the causal description of a system corresponds  an equivalent anticausal one, so that any observable $\mathbf{a}$ corresponds a conjugate ``anticausal observable'' $\mathbf{a}^\ast$. 
One must  have $(\mathbf{ab})^\ast =\mathbf{b}^\ast  \mathbf{a}^\ast $ (anticausality works in reverse order than causality) and $(\mathbf{a}^\ast)^\ast=\mathbf{a}$ (anti-anticausality$=$causality). Together with the compatibility assumption  $(\lambda \mathbf{a}+\mu \mathbf{b})^\ast = \lambda \mathbf{a}^\ast  + \mu \mathbf{b}^\ast $ (linearity), this means that the conjugation $^\ast$ is a real involution on $\mathcal{A}_{\scriptscriptstyle{\mathbb{R}}}$.

Causal reversibility puts strong constraints on the states, since
it implies that no choice of observable and state should  allow to distinguish the causal description from the anticausal one. 
Hence the states must be  $^\ast$\nobreakdash-symmetric real forms, such that $\varphi(\mathbf{a}^\ast )=\varphi(\mathbf{a})$, $\forall\,\mathbf{a}$.
The \emph{physical observables} (corresponding to physically measurable quantities) are the symmetric elements of $\mathcal{A}_{\scriptscriptstyle{\mathbb{R}}}$, such that $\mathbf{a}=\mathbf{a}^\ast$, while the anti-symmetric (or skew-symmetric) elements such that $\mathbf{a}=-\mathbf{a}^\ast$ are not physical (their expectation value is always zero) but have to be taken into account in order to have a consistent algebra structure for $\mathcal{A}_{\scriptscriptstyle{\mathbb{R}}}$.

Finally, for the states to give expectation values, so that $\varphi(\mathbf{a})$ has a probabilistic interpretation as in the classical case, the states $\varphi$ must  be the normalized \emph{positive} \emph{real} linear forms on $\mathcal{A}$, such that $\varphi(\mathbf{a}^\ast\mathbf{a})\ge 0$ (positivity), and 
$\varphi(\mathbf{1})=1$ (normalization). Indeed it turns out that  this standard positivity condition is equivalent (at least in the finite dimensional case \cite{David2011}), to the more physical but less stringent one $\varphi(\mathbf{a}^2)\ge 0$, for all $\mathbf{a}=\mathbf{a}^\ast$ (measuring a squared physical observable gives always a positive result).

Once this abstract definition of a real algebra of observables and of the corresponding states is obtained, I now show that, similarly to the known case of complex algebras, this algebra is a C$^\star$-algebra and can be represented as an algebra of operators on a real Hilbert space.
The theory of complex C$^\star$-algebra is very well known and the celebrated GNS construction can be used in this case to construct the Hilbert space out of pure states (see \cite{Haag96} and references therein), but there is a theory of real C$^\star$-algebra, (see \cite{Goodearl82} for definitions and basic results) that will be used here for a similar purpose.

Restricting $\mathcal{A}_{\scriptscriptstyle{\mathbb{R}}}$ to the bounded non-trival operators ($\mathbf{a}$ such that $\forall\varphi$, $\varphi(\mathbf{a}^\ast\mathbf{a})<\infty$ and $\exists \varphi : \varphi(\mathbf{a}^\ast\mathbf{a})>0$) and denoting $\mathcal{E}$ the corresponding (convex) set of  states (positive symmetric reals forms) on $\mathcal{A}_{\scriptscriptstyle{\mathbb{R}}}$,  
it is endowed with the norm
\begin{equation}
\label{ }
\| \mathbf{a}\| =\left(\sup\nolimits_{\varphi\in\mathcal{E}}\varphi(\mathbf{a}^\ast\mathbf{a}) \right)^{1/2}\ .
\end{equation}

In analogy with the well-known complex case, it is easy to show that this is indeed a norm and makes $\mathcal{A}_{\scriptscriptstyle{\mathbb{R}}}$
(or more exactly its completion $\overline{\mathcal{A}_{\scriptscriptstyle{\mathbb{R}}}}$) an abstract {C$^\star$-algebra}. 
Using the positivity of states and the Schwarz inequality one recovers,  by simple   arguments (similar to the complex case), the (in)equalities
$\|\lambda \mathbf{a}\|=|\lambda|\,\|\mathbf{a}\|$, $\|\mathbf{a}+\mathbf{b}\|\le \|\mathbf{a}\|+\|\mathbf{b}\|$, 
$\|\mathbf{ab}\|\le \|\mathbf{a}\|\,\|\mathbf{b}\|$, that make $\mathcal{A}_{\scriptscriptstyle{\mathbb{R}}}$ a \emph{real Banach algebra}, and the basic identity 
\begin{equation}
\label{CsCond}
\|\mathbf{a}^\ast \mathbf{a}\|=\|\mathbf{a}\|^2=\| \mathbf{a}^\ast\|^2
\end{equation}
(which is sufficient in the complex case to define an abstract {C$^\star$-algebra}).
One also has the additional non trivial condition
\begin{equation}
\label{realCsCond}
\mathbf{1}+\mathbf{a}^\ast\mathbf{a}\quad\text{is invertible}\quad\forall\, \mathbf{a}\in \mathcal{A}_{\scriptscriptstyle{\mathbb{R}}}
\end{equation}
Were (\ref{realCsCond})  false,
$\exists\; \mathbf{a}\; \&\; \mathbf{b}\neq 0 : \mathbf{b}+\mathbf{a}^\ast \mathbf{ab}=0$, hence
$\mathbf{b}^\ast\mathbf{b}+(\mathbf{ab})^\ast(\mathbf{ab})=0$, this contradicts positivity.
(\ref{CsCond}) and (\ref{realCsCond}) make $\mathcal{A}_{\scriptscriptstyle{\mathbb{R}}}$ 
a \emph{real} C$^\star$-algebra (see \cite{Goodearl82}), by ensuring that the  spectrum of the positive elements of $\mathcal{A}_{\scriptscriptstyle{\mathbb{R}}}$ is $\subset\mathbb{R}_+$.

I now use the mathematical results gathered in \cite{Goodearl82}. If $\mathcal{A}_{\scriptscriptstyle{\mathbb{R}}}$ is finite dimensional, purely algebraic methods (using the Wedderburn-Artin theorem) show that $\mathcal{A}_{\scriptscriptstyle{\mathbb{R}}}$ is a direct product of matrix algebras $M_n(K)$ over $K=\mathbb{R}$, $\mathbb{C}$ or $\mathbb{H}$ (the quaternions).
The physical observables $\mathbf{a}$ corresponds to the symmetric matrices $A$, and the states $\varphi$ corresponds to density matrices $\rho_\varphi$ (symmetric normalized $>0$ matrices such that $\varphi(\mathbf{a})=\mathrm{tr}(\rho_\varphi A)$ (this is nothing but the Born rule).
The infinite dimensional case is much more difficult (involving real analysis), but a theorem by Ingelstam \cite{Ingelstam64,Palmer70,Goodearl82} states that $\mathcal{A}_{\scriptscriptstyle{\mathbb{R}}}$ is indeed isomorphic to a 
symmetric 
closed sub-algebra of the algebra $B(\mathcal{H}_{\scriptscriptstyle{\mathbb{R}}})$ of bounded operators on some \emph{real} Hilbert space $\mathcal{H}_{\scriptscriptstyle{\mathbb{R}}}$. 
Thus the announced result is obtained: starting from a real algebra of observables, and using causality and reversibility, this algebra must be represented as an algebra of operators on a real Hilbert space.

Let me just recall  the reason for  using complex algebras and complex Hilbert spaces in quantum physics (complex C$^*$-algebras are a special case of real algebras, real states and complex states being in correspondance).
Quantum theories based solely on real Hilbert spaces has been investigated, notably by Stueckelberg \cite{Stueckelberg1960}, but
problems occur when trying to construct relativistic theories, and no satisfactory solutions seems to exist.
Problems are also known to occur in quaternionic quantum mechanics \cite{FinkelsteinEtAl1962}.
This can be understood by taking into account the physical requirement of  locality and of separability. For instance (see \cite{Wootters1990}) consider two causally independent subsystems $\mathcal{S}_1$ and $\mathcal{S}_2$, forming a composite system $\mathcal{S}=\mathcal{S}_1\cup\mathcal{S}_2$, with respective operator algebras $\mathcal{A}_1$, $\mathcal{A}_2$ and $\mathcal{A}=\mathcal{A}_1\otimes\mathcal{A}_2$. In the complex case the \emph{physical observables} $\mathbf{a}$ for $\mathcal{S}$ (the symmetric operators) can always be built as linear combinations of independent products of \emph{physical observables} of the two subsystems $\mathbf{a}=\sum_i\mathbf{a}_1^{\scriptscriptstyle{(i)}}\otimes \mathbf{a}_2^{\scriptscriptstyle{(i)}}$. This is not the case for real algebras, which leads to a clash with  locality and separability.

I now briefly discuss the role of reversibility in the quantum logic (or quantum propositional calculus) formalism. 
This formalism, initiated in \cite{BirkVNeumann36}, pursued in \cite{Mackey63,Jauch68,Piron64},
aims at defining the abstract analog of the set of projectors in a Hilbert space (corresponding to ideal projective measurements), but without assuming any predefined C$^*$-algebra structure. There are many variants of this approach. I shall rely on the classical review \cite{BeltCassi81}. Contrary to the first part of this letter, 
I shall not present here a new version of the formalism, but I shall discuss in a seemingly new way two points where causal reversibility and locality enter in the existing formalism. A complete discussion is left to \cite{David2011}.

The central concepts are the propositions/tests (the abstract analog of
 projectors 
in the algebraic formalism), and the states  (the different possible knowledge/expectations one can have on a system).
The propositions are operations on the system $\mathcal{S}$ that return the boolean value TRUE or FALSE ($1$ or $0$). 
This is in general a non-deterministic process, and  depends on the initial state $\varphi$ of $\mathcal{S}$. 
After the operation, $\mathcal{S}$ needs not to be in the initial  state $\varphi$ (since we have in general gained some information on $\mathcal{S}$). The structure of the states and observables for $\mathcal{S}$ is constructed out of five axioms. The first two are  related to causality.

\noindent  (I) \emph{Order relation and causality} : There is a logical relation $\preceq$  between propositions, 
making the set of propositions $\mathcal{L}$ a POSET. 
Amongst the various way to define $a\preceq b$, I choose the causal relation:
``starting from any initial state $\varphi$ and asking $a$, then $b$, if $a$ is found true, then $b$ will be found true''.
I recall that making  $\preceq$ a partial order relation means that once $a$ has been found to be true, then applying any $b$ such that $a\preceq b$ does not change the state (ideal measurements).

\noindent (II) \emph{AND and OR} :  The logical cunjunction $\wedge$ (the {unique greatest proposition} $a\wedge b$ such that
$a\wedge b\preceq a$ and $b$) and join $\vee$  (the {unique smallest proposition} $a\vee b$ such that
$a$ and $b\preceq a\vee b$) exist, and make $\mathcal{L}$ a \emph{complete lattice} (CL).
See \cite{Jauch68,Piron64} for  justifications.

\noindent (III) \emph{NOT and Reversibility} : 
It is the third axiom that I discuss here in a new way.
The order relation $\preceq$ is clearly causal (``if...then...''). When discussing what happens if  a proposition $a$ is found false, one has to define its \emph{negation} or \emph{complement} $\neg a$. 
It is at this crucial point that the concept of reversibility  enters in the formalism. 
Indeed one can define $\neg a$ as ``if $a$ is found true, then $\neg a$ will be found false'',
or alternatively as  ``if $\neg a$ is found true, then $a$ will be found false''.
But with {(I)} and {(II)} only these two definitions are \emph{not equivalent}! 
A third axiom is to assume that they are indeed equivalent, i.e.
\begin{equation}
\label{NegRev1}
a \preceq b\ \iff  \neg b\preceq \neg a
\end{equation}
(as in classical logic).
It also means that the $a\preceq b$ relation, defined in {(I)} as ``if $a$ is found true, then $b$ \emph{will be} found true'', is equivalently defined by ``if $a$ is found true, then $b$ \emph{was} found true''. This means that
the causal order structure on propositions is in fact independent of the choice of a causal arrow.
Therefore (\ref{NegRev1}) amounts to  \emph{the assumption of causal reversibility}.
This is our important point which, albeit simple, does not seem to have been discussed before.
(\ref{NegRev1}) also implies that one should expect a complete symmetry between causal prediction and causal retrodiction (a well known a posteriori property of the standard quantum formalism 
\cite{Aharonov1964}).

$\mathcal{L}$ is now an \emph{orthocomplemented complete lattice} (OCL).
Two propositions are said to be orthogonal and noted $a\perp b$ iff $a\preceq \neg b$.

I do not discuss the remaining axioms.
\noindent (IV) \emph{Weak modularity} %
implies that the orthocomplementation is unique, and is crucial for the formalism to describe consistently the subsystems and the subsets of states of some larger system.
\noindent (V) \emph{AC and Minimal Propositions} :
Finally one assumes that the lattice $\mathcal{L}$ is atomic (A), i.e. can be constructed out of  ``minimal propositions'' (atoms) that, when true, specify completely the state of the system (the analog of projection on pure states), and satisfy the covering property (C) which means (crudely) that any test $b$ on a pure state gives a pure state (this is also related to reversiblity).
NB: The need of minimal propositions limits the formalism to quantum systems described by Type I von Neumann algebras, but this encompasses some infinite systems like the vacuum sector of extended quantum systems.

The quantum logic formalism leads to a convincing derivation of the algebraic formulation of quantum theories.
A set of mathematical theorems 
(see \cite{BeltCassi81}) states that any complete (irreducible) orthomodular AC lattice can be represented 
({except for the analog of the 2 and 3 states systems})
as the lattice of the closed subspaces of a left-module  $V$ (generalization of  vector spaces) on a division ring $K$  such that
(i)  $K$ has an involution $^\star$; 
(ii)  $V$ has a non degenerate Hermitian form $f:\ V\times V \to K$;
(iii)  $f$ defines an orthogonal projection and associates to each linear subspace $M$ of $V$ its orthogonal $M^\perp$, closed subspaces being the $M$ such that $(M^\perp)^\perp=M$;
(iv) $f$ is orthomodular, i.e. for any closed $M$, $M^\perp+M=V$.
In particular, the structure $(\preceq\, ,\, \wedge\, ,\, \neg)$ on the lattice $\mathcal{L}$ is isomorphic to the standard structure $(\subseteq\ ,\,\cap \,,\,\perp)$
in the space $\mathcal{L}(V)$ of closed linear subspaces of $V$.
There is some  $a\in V$ with ``norm''  unity  $f(a,a)=1$.
The propositions of $\mathcal{L}$ can thus be identified with the projections on the closed subspaces of $V$.
\noindent\emph{Locality and Sol\`er's theorem}:
The restriction to the standard rings $K=\mathbb{R}$, $\mathbb{C}$ or $\mathbb{H}$ is mathematically obtained through Soler's theorem \cite{Soler95} (and its extensions). 
It states that for $\mathcal{L}=\mathcal{L}_K(V)$ an irreducible OM AC lattice, as discussed above,
if there is an infinite family of orthonormal (and in fact orthogonal) vectors in $V$ (i.e. an infinite set of mutually orthogonal atoms in $\mathcal{L}$) then $K$ can only be $\mathbb{R}$, $\mathbb{C}$ or $\mathbb{H}$.

Combining this theorem with the assumption of locality seems in fact sufficient. Here is a heuristic but physical argument, seemingly new.
Let me consider the case where the physical space in which the system is defined to be infinite (flat) space or some regular lattice, so that it can be separated into causally independent pieces $\mathcal{O}_\alpha$ (labelled by $\alpha\in\Lambda$ some infinite lattice). It is sufficient to have one single proposition $a_\alpha$ relative to each $\mathcal{O}_\alpha$ only (for instance ``there is one particle in $\mathcal{O}_\alpha$'') to build an infinite family of mutually orthogonal propositions $b_\alpha=a_\alpha\wedge(\bigwedge_{\beta\neq\alpha}\neg a_\beta)$ in $\mathcal{L}$. 
Out of the $b_\alpha$, thanks to the atomic property (A), we can extract an infinite family of orthogonal atoms $c_\alpha$, i.e. of orthogonal vectors in $V$. Q.E.D.
The quantum information inclined reader (not afraid of infinite dimensional Hilbert spaces) may prefer to apply a similar argument to the infinite family of causally independent subsystems ...((system+ancilla)+Ancilla)+...
Recovering the standard representation in terms of complex Hilbert spaces is done by invoking separability as above, Gleason's theorem and reconstruction of general observables out of the projectors.

To conclude, in this letter I have separated the respective role of causality, reversibility and locality in the quantum formalism, stressing in a new way the importance of causal reversibility. The discussion is made for the algebraic  formalism, and for a simple version of the quantum logic formalism.
I chose a somewhat traditional point of view, and did not  discuss  the  approaches where the quantum theory is formulated through purely informational principles. It would be interesting to see if here too a specific role for reversibility can be singled out (reversibility is discussed  in e.g.  \cite{CoeckeLal2010} and references therein).
Discussing these issues for quantum gravity is beyond the scope of this letter.
But a fundamental question raised by this discussion is whether, in any reasonable pre-quantum (or post-quantum...) theory 
where the ordinary notion of causality is  somehow modified, reversibility, considered as a separate concept, has to be kept as a fundamental principle, or is just an ``emergent'' property of the quantum world as we experience it.

\end{document}